\documentclass[12pt,preprint]{aastex}

\shorttitle{Occultation by a Planetesimal }
\shortauthors{Chen et al.}

\begin{document}

\title{
A Possible Detection of 
Occultation by a Proto-planetary Clump in GM\,Cephei
    }

\author{
W. P. Chen\altaffilmark{1}, S. C.-L. Hu\altaffilmark{1,2}, R. Errmann\altaffilmark{3},
Ch. Adam\altaffilmark{3},
S. Baar\altaffilmark{3},
A. Berndt\altaffilmark{3},
L. Bukowiecki\altaffilmark{4},
D. P. Dimitrov\altaffilmark{5}
T. Eisenbei\ss\altaffilmark{3},
S. Fiedler\altaffilmark{3},
Ch. Ginski\altaffilmark{3},
C. Gr\"afe\altaffilmark{3,6}, 
J. K. Guo\altaffilmark{1}
M. M. Hohle\altaffilmark{3},
H. Y. Hsiao\altaffilmark{1}, 
R. Janulis\altaffilmark{7},
M. Kitze\altaffilmark{3},
H. C. Lin\altaffilmark{1}, 
C. S. Lin\altaffilmark{1}, 
G. Maciejewski\altaffilmark{3,4},
C. Marka\altaffilmark{3},
L. Marschall\altaffilmark{8},
M. Moualla\altaffilmark{3,9},
M. Mugrauer\altaffilmark{3},
R. Neuh\"auser\altaffilmark{3},
T. Pribulla\altaffilmark{3,10},
St. Raetz\altaffilmark{3},
T. R\"oll\altaffilmark{3},
E. Schmidt\altaffilmark{3},
J. Schmidt\altaffilmark{3},
T. O. B. Schmidt\altaffilmark{3},
M. Seeliger\altaffilmark{3},
L. Trepl\altaffilmark{3},
C. Brice\~{n}o\altaffilmark{11}, 
R. Chini\altaffilmark{12},   
E.L.N. Jensen\altaffilmark{13},
E. H. Nikogossian\altaffilmark{14} 
A. K. Pandey\altaffilmark{15}, 
J. Sperauskas\altaffilmark{7},
H. Takahashi\altaffilmark{16},
F. M. Walter\altaffilmark{17},
Z.-Y. Wu\altaffilmark{18},
X. Zhou\altaffilmark{18}
}

\altaffiltext{1}{Graduate Institute of Astronomy, National Central University, 300 Jhongda Road, Jhongli 32001, Taiwan}
\altaffiltext{2}{Taipei Astronomical Museum, 363 Jihe Rd., Shilin, Taipei 11160, Taiwan}
\altaffiltext{3}{Astrophysikalisches Institut und Universit\"ats-Sternwarte, FSU Jena,
                    Schillerg\"a\ss chen 2-3, D-07745 Jena, Germany}
\altaffiltext{4}{Toru\'{n} Centre for Astronomy, Nicolaus Copernicus University, Gagarina 11, PL87-100 Toru\'{n}, Poland}
\altaffiltext{5}{Institute of Astronomy and NAO, Bulg. Acad. Sc., 72 Tsarigradsko Chaussee Blvd., 1784 Sofia, Bulgaria}
\altaffiltext{6}{Christian-Albrechts-Universit\"at Kiel, Leibnizstra\ss e 15, D-24098 Kiel, Germany}
\altaffiltext{7}{Moletai Observatory, Vilnius University, Lithuania}
\altaffiltext{8}{Gettysburg College Observatory, Department of Physics, 300 North Washington St., Gettysburg, PA 17325, USA}
\altaffiltext{9}{Now at Tishreen University in Lattakia, Syria}
\altaffiltext{10}{Astronomical Institute, Slovak Academy of Sciences, 059 60, Tatransk\'a Lomnica, Slovakia}
\altaffiltext{11}{Centro de Investigaciones de Astronomia, Apartado Postal 264, Merida 5101, Venezuela}
\altaffiltext{12}{Instituto de Astronom\'{i}a, Universidad Cat\'{o}lica del Norte, Antofagasta, Chile}
\altaffiltext{13}{Dept. of Physics and Astronomy, Swarthmore College, Swarthmore, PA 19081-1390, USA}
\altaffiltext{14}{Byurakan Astrophysical Observatory, 378433 Byurakan, Armenia}
\altaffiltext{15}{Aryabhatta Research Institute of Observational Science, Manora Peak, Naini Tal, 263 129, Uttarakhand, India}
\altaffiltext{16}{Institute of Astronomy, The University of Tokyo, 2-21-1 Osawa, Mitaka, Tokyo, 181-0015, Japan}
\altaffiltext{17}{Department of Physics and Astronomy, Stony Brook University, Stony Brook, NY 11794-3800, USA}
\altaffiltext{18}{Key Laboratory of Optical Astronomy, NAO, Chinese Academy of Sciences, 20A Datun Road, Beijing 100012, China}


\begin{abstract}
GM\,Cep in the young ($\sim4$~Myr) open cluster Trumpler~37 has been known to be an abrupt 
variable and to have a circumstellar disk with very active accretion.  Our monitoring
observations in 2009--2011 revealed the star to show sporadic flare events, each with
brightening of $\lesssim0.5$~mag lasting for days.  These brightening events, associated with a color change 
toward the blue, should originate from an increased accretion
activity.  Moreover, the star also underwent a brightness drop of $\sim1$~mag lasting for
about a month, during which the star became bluer when fainter.  Such brightness drops seem to have 
a recurrence time scale of a year, as evidenced in our data and the photometric behavior of GM\,Cep  
over a century.  Between consecutive drops, the star brightened gradually by about 1~mag and became blue 
at peak luminosity.  We propose that the drop is caused by obscuration of the central star by an orbiting dust 
concentration.  The UX Orionis type of activity in GM\,Cep therefore exemplifies the disk inhomogeneity process 
in transition between grain coagulation and planetesimal formation in a young
circumstellar disk.
\end{abstract}

\keywords{ Occultations --- Planets and satellites: formation --- Protoplanetary disks --- 
           Stars: Individual: GM Cep --- Stars: pre-main sequence --- Stars: variables: T Tauri, Herbig Ae/Be
        }

\section{Introduction}

The current paradigm suggests that stars are formed in dense molecular cores, and planets
are formed, almost contemporaneously with the star, in circumstellar disks.  The grain
growth process already initiated in the parental molecular cloud continues to
produce progressively larger solid bodies.  Details are still lacking in how grain coagulation
proceeds to eventual planet formation in a turbulent disk.  Competing theories include 
gravitational instability \citep{saf72,gol73,joh07} and planetesimal accretion \citep{wei00}.
In any case, density inhomogeneities in the young stellar disk mark  
the critical first step in the process.  
Measurements of the fraction of stars with infrared excess---arising from thermal emission by circumstellar
dust---indicates a clearing time scale of optically thick disks in less than $\sim10$~Myr \citep{mam04,bri07,hil08}.  
Observationally, this epoch corresponds to pre-main sequence (PMS) stellar evolution
from disk-bearing classical T Tauri stars (CTTSs) to weak-lined T Tauri stars 
with no optically thick disks.

The open cluster \object{Trumpler 37} (Tr\,37), at a heliocentric distance of 870~pc \citep{con02}, 
is associated with the prominent \ion{H}{2} region IC\,1396, and is a part of the Cepheus 
OB2 association.  With a disk frequency of $\sim39$\% \citep{mer09}, and an age of
1--4~Myr \citep{mar90,pat95,sic05}, Tr\,37 serves as a good target to search for and to characterize 
exoplanets in formation and early evolutionary stages \citep[see][and references therein on Tr\,37]{neu11}.  

\object{GM Cep} (RA = 21:38:17.3, Dec = +57:31:23,
J2000) is a solar type variable in Tr\,37.  The star has a spectral type of G7 to K0, an
estimated mass of 2.1~M$_\sun$ and a radius of 3--6~R$_\sun$ \citep{sic08}.
The youth of GM\,Cep is exemplified by its emission-line spectrum, prominent infrared
excess \citep{sic08}, and X-ray emission \citep{mer09}, all characteristics of a CTTS.
The star has a circumstellar disk \citep{mer09}, with an accretion rate up to
$10^{-6}$~M$_\sun$~yr$^{-1}$, which is 2--3 orders higher than the median value of the
CTTSs in Tr\,37 \citep{sic06}.  It is also one of the fastest rotators in the cluster,
with $v \sin i \sim43.2$~km~s$^{-1}$ \citep{sic08}.

Most PMS objects are variables.  \citet{her94} classified such variability into
three categories.  One class of variation is caused by rotational modulation of cool star spots.
Another class of variation arises because of unsteady accretion onto a hot spot on the stellar surface; stars
of this type are called EXors, with EX Lupi being the most extreme case.  
Stars with the third kind of variation, 
called UX Orionis type variables or UXors, are those which experience variable obscuration by circumstellar dust
clumps.  About a dozen UXors have been identified so far, with some showing cyclic variability with periods 
ranging from 8.2~days \citep{bou03} to 11.2~years \citep{gri98}.  

GM\,Cep is known to be an abrupt variable, but interpretations on its variability have been
controversial.  \citet{sic08} collected photometry of the star from 1952 to 2007
in the literature, supplementing with their own intensive multi-wavelength observations, and
suggested GM\,Cep to be an EXor type variable, i.e., with outbursts and
accretion flares.  \citet{xia10} measured archival plates taken between 1895 and 1993,
and concluded otherwise---that the variability in the century-long light curve is dominated by
dips (possibly from extinction) superposed on quiescent states.  If this is the case, GM\,Cep should
be a UXor type variable, as claimed also by \citet{sem11}.

GM\,Cep has been observed by the Young Exoplanet Transit Initiative (YETI) collaboration, a network
of small telescopes in different longitude zones \citep{neu11}.  In addition to the YETI data, the observations 
reported here also included those collected during non-YETI campaign time, by the SLT 40-cm telescope 
at Lulin in Taiwan, the Tenagra~II 81-cm telescope, in Arizona, USA, the Jena University Observatory 
25~cm and 90/60~cm telescopes in Germany, and the 1.5~m telescope of Moleitai Observatory in Lithuania.  
For the list of YETI telescope and instrument parameters, please refer to \citet{neu11}.  While the 
primary goal of the YETI campaigns, each with uninterrupted monitoring of a target cluster for 7--10 days, 
is to search for exoplanet transit events in young open clusters---hence finding possibly the youngest 
exoplanets---the continuous and high-cadence observations produce data sets also valuable for 
young stellar variability study much relevant to planet formation \citep{bou03}.
Here we present the light curve of GM\,Cep from 2009 to 2011 that reveals T~Tauri-type flares and 
UXor-type variability, with the possible detection of cyclic occultation events by a dust clump in 
the circumstellar disk.

\section{Light Curves and Color Variations }

\begin{figure}
   \includegraphics[angle=90,width=1.0\textwidth]{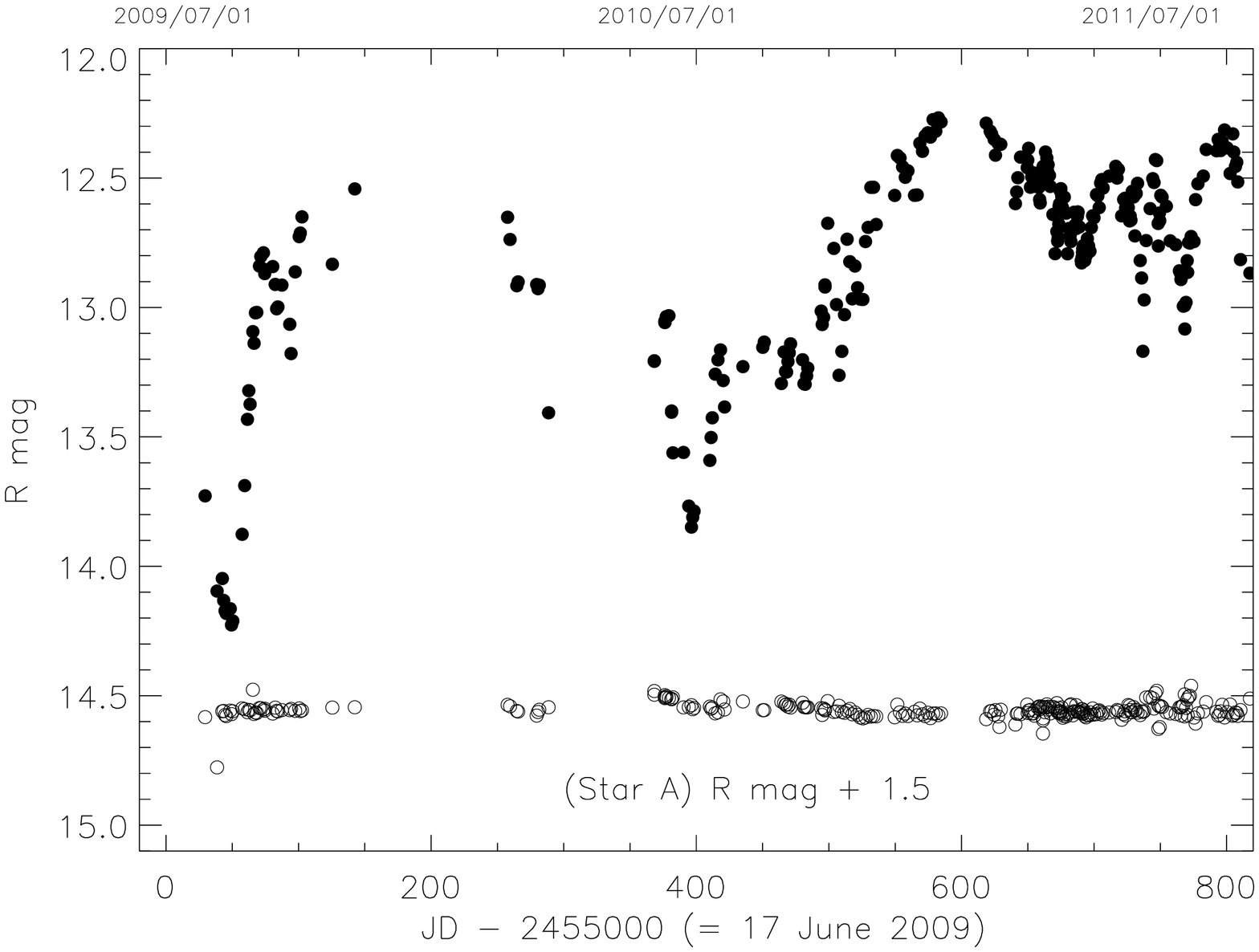}
   \caption{The R-band light curves of GM\,Cep (top) and of a comparison star (bottom, offset by 1.5~mag for display clarity) from
           mid-2009 to mid-2011.  Typical photometric errors (0.005~mag) are smaller than the sizes of the symbols and are not shown.
            } 
   \label{fig:01}
\end{figure}

All the CCD images were processed by the standard procedure of bias, dark and flatfield correction.  
The photometry of GM\,Cep was calibrated by a linear regression with the seven comparison stars 
listed by \citet{xia10}.  Images taken under inferior sky conditions were excluded in the analysis. 
Figure~\ref{fig:01} shows the light curves of GM\,Cep and one of the comparison stars observed from mid-2009 to mid-2011.
The variability of GM\,Cep is obvious.  The star experienced a sharp brightening soon after our 
observations started in mid-2009, prompting us to follow this star closely beyond the YETI campaigns. 
Our intense monitoring started in 2010.  A brightness dip, with a depth of $\Delta R \sim0.82$~mag, lasting for 39~days, 
occurred, followed by gradual brightening (by $\sim1$~mag) and fading.  
The falling and rising parts of the dip are roughly symmetric.   In 2011, a dip also happened, but with 
rapid fluctuations.   The star fluctuated some $\Delta R\sim1.7$~mag in 2010 and also in 2009.  We conclude that the 
sharp brightening in 2009 corresponded to the rising part of the dip seen in 2010.  If so, the recurrent time 
scale of the dip would be 346~days, and the minimum of the dip brightened from 2009 ($R\sim14.2$~mag), 2010 ($R\sim13.9$~mag), 
to 2011 ($R\sim13.2$~mag).  When this trend is taken out, the gradual brightening and fading is more or less symmetric 
in time with the peak luminosity happening between two consecutive dips, much like the round-topped light curves seen in 
contact binaries.  Such repeated dips plus a slow brightening and 
fading can be seen in the long-term light curve reported by \citet{xia10}, who claimed no periodicity in the data perhaps 
because of the sparse sampling of the data.  

\begin{figure}
   \includegraphics[angle=90,width=1.0\textwidth]{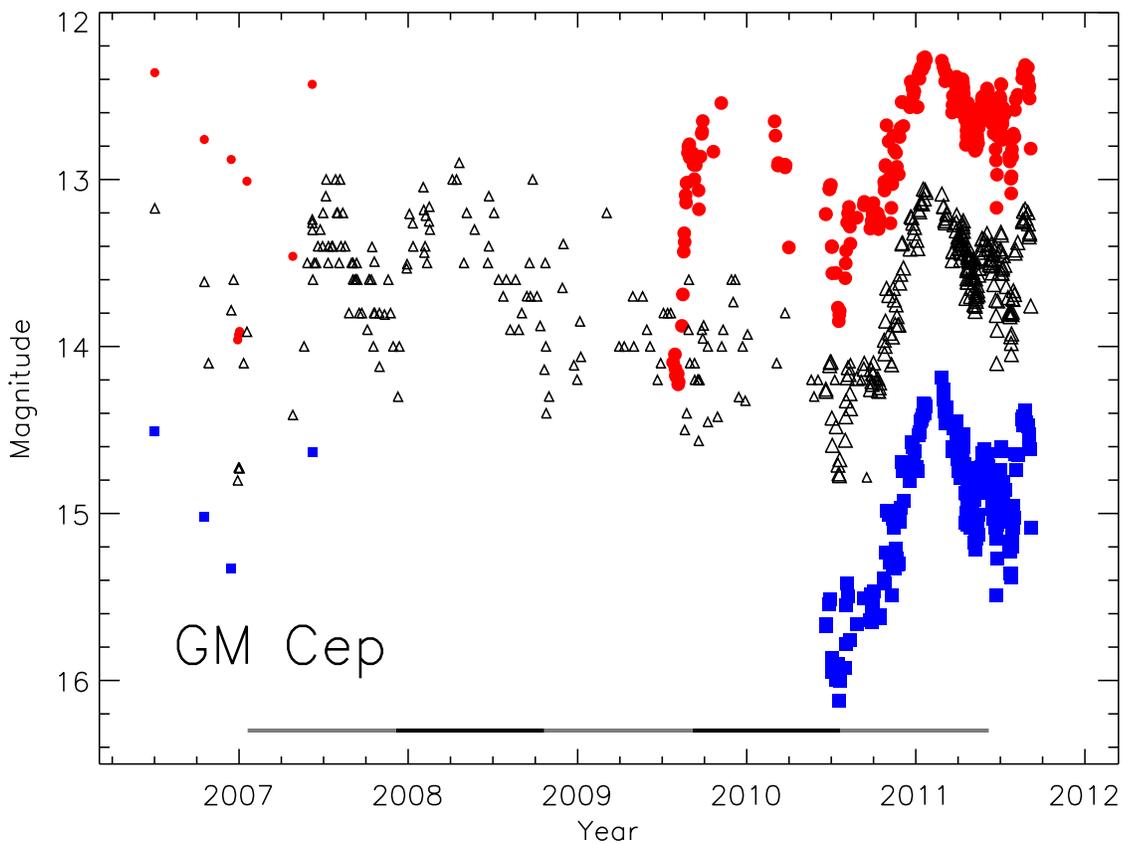}
  \caption{Light curves of GM\,Cep in $B$ (denoted by red circles), $V$ (black triangles), and $R$ (blue squares) bands
              between late 2006 to 2011.  Symbols with larger sizes, i.e., those after 2009, represent our observations.  
              Each segment of the horizontal black and gray line is shown for duration of 320~days, to coincide roughly 
              with the brightness dips.
    }
\label{fig:02}
\end{figure}

\begin{figure}
   \includegraphics[height=\textheight]{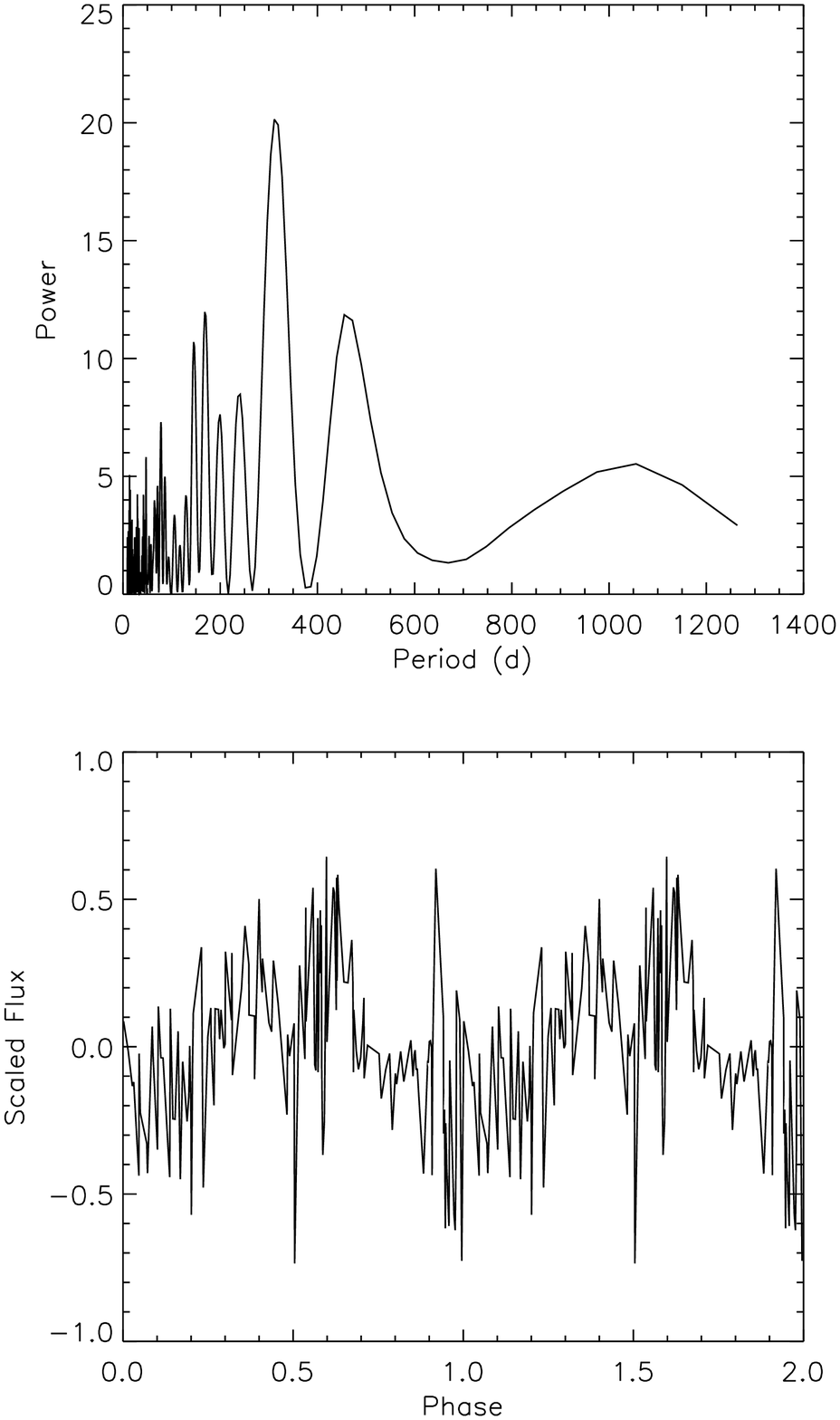}
  \caption{({\it Top}) Power spectrum of the light curve in Figure~\ref{fig:02} analyzed by the Lomb-Scargle algorithm, peaking 
           at the period of 311~d. ({\it Bottom}) Phased light curve with the 311~d period. 
    }
\label{fig:03}
\end{figure}

Figure~\ref{fig:02} shows the light curves of GM\,Cep in $B, V, R$ bands since late 2006, 
with additional data taken from \citet{sic08} and AAVSO.  Analysis by the NStED (NASA/IPAC/NExScI Star and 
Exoplanet Database) Periodogram Service, based on the Lomb-Scargle algorithm, shows the first-ranked peroid to be 
311~d with a broad peak in the power spectrum, suggesting a quasi-periodicity, as shown in Figure~\ref{fig:03}.  Such a recurrence time 
scale of 310--320~d indeed seems to coincide with the minima in the light curve (see Figure~\ref{fig:02}) at least 
for the last 5 cycles for which sampling has been sufficiently dense \citep{hu12}.  
In addition, superimposed on the above light variations, there are sporadic flaring-like episodes with amplitude less 
than 0.5~mag, each lasting for about 10~d, characteristic of T~Tauri activity.

\begin{figure}
\includegraphics[angle=90,width=1.0\textwidth]{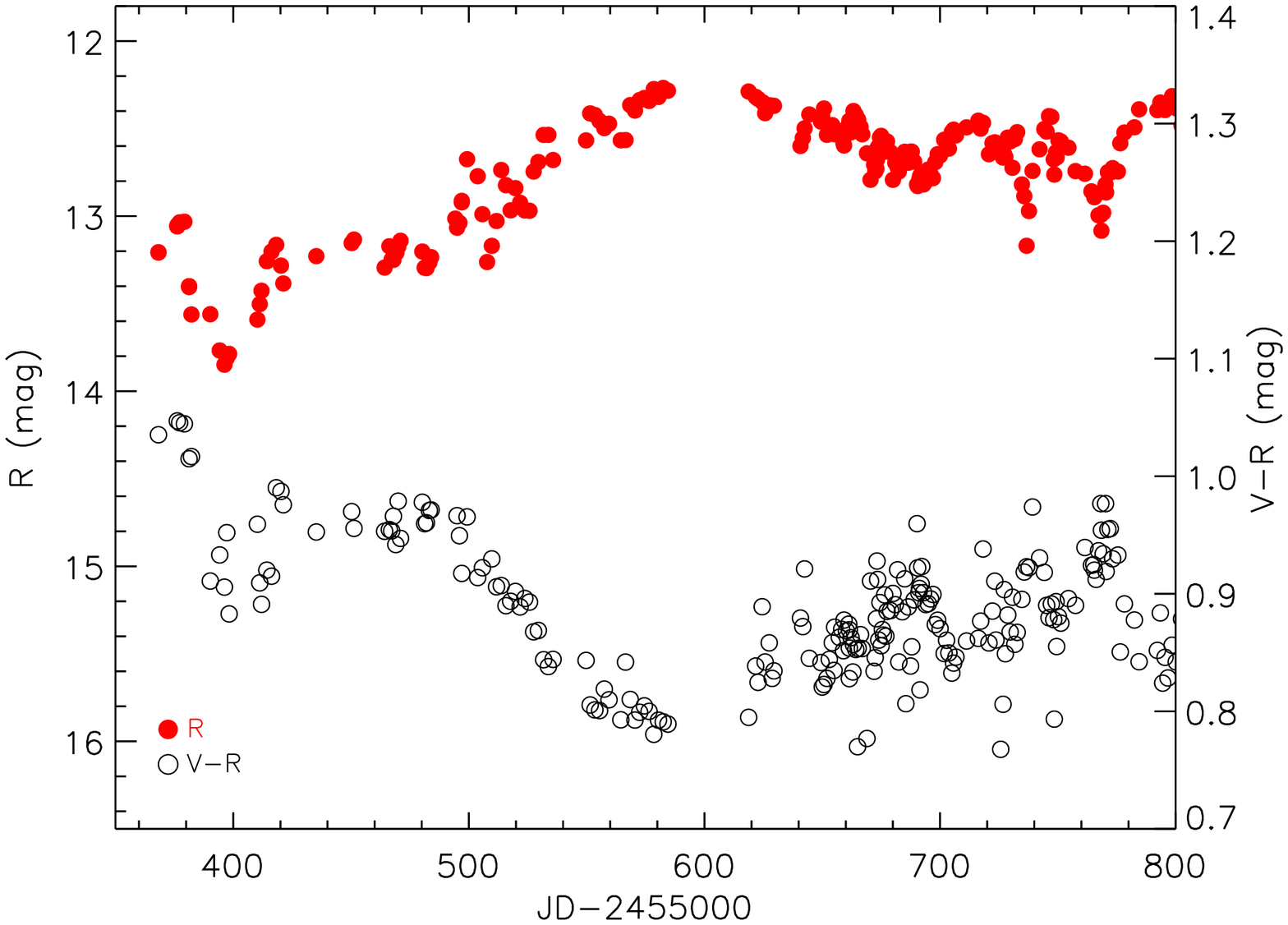}
   \caption{
          The $R$-band light curve (left y-axis) and the $V-R$ color variations (right y-axis, redder to the top)
          of GM\,Cep from mid-2010 to mid-2011.  Note the star became blue when faintest and brightest.
             }
  \label{fig:04}
\end{figure}

While the YETI campaigns are carried out in the $R$ band, our intensified observations of GM\,Cep since 2010 included also 
those taken in the $V$ band.  The color changes during the dip, as well as during the brightening and fading episodes, are 
particularly revealing.  Fig.~\ref{fig:04} shows the $R$-band light curve and $V-R$ color variations in 2010/2011.  
The dip in the beginning has a depth of about $\Delta V\sim0.68$~mag; so while the star became fainter 
(depth in $R$ was 0.82~mag), the $V-R$ value decreased, i.e., its color turned bluer.  During the general brightening, 
the star also became bluer.  
  
To summarize, the light curve of GM\,Cep is characterized by (1)~a brightness dip of about 1~mag lasting for a month,
with a recurrence time scale of about a year, (2)~in between the dips, a gradual brightening of about 1~mag, followed by
a roughly symmetric fading, and superimposed on the above two, (3)~intermittent flares $\lesssim0.5$~mag, each lasting 
for several days.  

\section{Discussion}

The abrupt behavior in GM\,Cep's light curve is not uncommon among Herbig Ae/Be stars, with modulations 
of various time scales, i.e., "cyclic but not exactly periodic" \citep{her99}, superimposed on deep minima.  
A flare with a blue color can be accounted for by enhanced accretion of clumpy material.  \citet{sem11} published the 
$B, V, R, I$ light curves of GM\,Cep covering from mid-2008, i.e., one year earlier, 
but in lower cadence, than our data.  Their data showed $R\sim12$~mag in 2008 with no obvious dips, an obscuration 
event in 2009, and another one in 2010.  These authors proposed that GM\,Cep is a UXor variable.  
At the end of their observations, in early 2011, the star reached again $R\sim12$~mag, shown also in our data.  

The most striking feature of the light curve of GM\,Cep is the month-long dips.  
There are various possible mechanisms to produce such a phenomenon, e.g., by star spots or a rotating accretion column, 
which has a typical time scale of a few hours to days.  A notable case, the T~Tauri star AA\,Tau, is known to 
show deep fading ($\sim 1.4$~mag) lasting for about a week, believed to be caused by occultation by a warp 
in the magnetospheric accretion disk \citep{bou99}, with a quasi-cyclic time scale of 8.2~d \citep{bou03,bou07}.  The dip 
phenomenon appears to be common among young stars with inner dusty disks \citep{her99}.  In a study by CoRot 
satellite of the young star cluster NGC\,2264, \citet{ale10} found a fraction of 30--40\% young stars exihibiting 
obscuration variations.      

We propose that the month-long dip seen in GM\,Cep is a manifestation of obscuration by an orbiting dust concentration 
in the circumstellar disk, i.e., GM\,Cep is a UXor-type variable, as reported by \citet{xia10} and by \citet{sem11}.  
If so, the orbital period of the dip gives information on the distance of the clump from the star, whereas the 
duration of the obscuration and amount of starlight extinction, give, respectively, the size and the column density of the clump.
The mass of the star is uncertain for this PMS star, but assuming 2.1~M$_\sun$ \citep {sic08}, a Keplerian motion,
and a period of $P$=320~days, the orbital distance of the clump would be $r\sim1.2$~AU.  The duration of the 
obscuration $t\sim39$~days is related to the half-size of the clump $R_c$ by $t/P =(2R_c)/(2\pi r)$; hence $R_c\sim0.4$~AU, or about
15--30 stellar radii \citep{sic08}.  

The extinction $A_{\lambda}$ at wavelength $\lambda$ is related to the amount of obscuring dust along the line
of sight, i.e., $A_{\lambda}=1.086 \, N_d \, \sigma_d \, Q_{ext}$, where $N_d$ is the column density of the dust grains,
$\sigma_d$ is the geometric cross section of a grain of a radius of $a$ $\sigma_d=\pi a^2$, and $Q_{ext}$ is the 
dimensionless extinction efficiency factor.  Stars as young as GM\,Cep should have large grains settled into 
the midplane, but because the disk is inclined \citep{sic08}, we assume that the obscuration is caused 
mostly by small dust grains with an average radius of $a\sim0.1~\mu$m, thus $Q_{ext}\sim 1$, cautiously noting the possibility 
of abnormal dust sizes in the disk \citep{sic08}.  It follows from the observed obscuration of $0.68$~mag in the $V$ band   
that $ N_d = 2.0 \times 10^9 \,\mbox{cm}^{-2}$.  This amount of intervening dust is hardly excessive.  The flux drop during 
the dip phase, $\sim1$~mag, is comparable to the extinction of the star $A_V\sim1.5$ \citep{con02,sic04}, a 
value commonly seen among CTTSs.  The moderate extinction also indicates a line of sight out of the disk plane. 
What is intriguing in GM\,Cep of course 
is the distinct on-off behavior of the obscuration.  The column mass density is, given the same amount of extinction, 
proportional to the dust size $a$, and in this case is $\Sigma\sim2.9\times10^{-5}\, \mbox{g~cm}^{-2}$.  Even for 
$a=10~\mu\mbox{m}$ grains, the column mass density would be still several orders less than the minimum solar nebula, for which 
$\Sigma$ is a few thousands $\mbox{g~cm}^{-2}$ at 1~AU \citep{wei77}).

It is not clear whether the clump has a line-of-sight (radial) dimension comparable to its transverse size ($2R_c$) 
or is merely a ringlet.  Even if it is spherical, thus yielding the maximum mass, the mean volume density would be 
$ n_d = N_d / 2R_c = 1.7 \times 10^{-4}\, \mbox {cm}^{-3}$ at the clump's center.  Given the proximity of the clump to 
the star ($r=1.2$~AU), we assume the dust composition to be mostly silicates, having an average density of 
$ \rho=3.5~$g\,cm$^{-3}$.  This leads to an estimated mass of $ M_d= 2.3 \times 10^{21}\,\mbox{g}$ for the clump, 
about that of an asteroid, if the mass is uniformly spread.  For a clump this substantial 
in size, our line of sight does not need to line up to the orbital plane in order to detect the occultation.  
From the fast rotation, the infrared spectral energy distribution, and the H$\alpha$ profile, an intermediate inclination 
angle was inferred \citep{sic08}.  A clump extending in radial direction  
would have been tidally unstable.  The clump is thus extended along the orbit, but short radially.   

The blueing phenomenon during the obscuration is most puzzling.  It has been seen in UX\,Ori itself \citep{her99} and other 
UXors \citep{gri01}.    
\citet{sem11} reported also the ``color reversal'' or the blueing effect in GM\,Cep, and attributed it to possible 
anomalous dust properties, or disk geometry such as self-shadowing or a piled up wall in the inner disk \citep{dul03}.  
One appealing proposal by \citet{gri94} is that blueing happens when dust along the line of sight completely dims the star, and   
dust particles near the line of sight scatter preferentially blue light into the view, a mechanism supported 
by increased polarization during maximum extinction.  In GM\,Cep when the clump blocks out the star, either the hot boundary layer ---  
a region between the star and the active accretion disk --- or the magnetospheric accetion column, must have  contributed 
much to the emission during the dip phase.
 

It is interesting to note that, except for the flare events, the light curve of GM\,Cep, namely repeated occultation modulated 
by gradual, symmetric brightening and fading, bears resemblance to that of an eclipsing binary or an exoplanet transit with 
phase variations \citep{bor09}, though the time and flux change scales are vastly different.  In GM\,Cep the flares
are caused by enhanced accretion activity, and the dip, as we propose here, by occultation of the central star by a patch of
dust in the circumstellar disk.  The gradual brightening and fading, then, is the result of the orbital modulation of  
reflected starlight, as witnessed in high-precision light curves of eclipsing binaries or transiting exoplanets \citep{bor09}.  
Without the shape information of the clump, it is difficult to quantify this effect.  But the amount of reflected 
light allows us to estimate the height of the clump.  If the yearly brightening trend in 2009--2011 is removed, the gradual brightening 
in 2010 amounted to $\sim0.7$~mag, meaning approximately an equal contribution between the reflected light and the direct 
starlight.  Without knowledge on the density distribution and optical properties
of dust, we made a simple analogy of dust grains as a translucent mirror, made up of a total number of $N_{\rm tot}$ 
particles.  Assuming the Bond albedo $a_B$, the reflected light is $(L_*/4\pi r^2)\; \pi a^2 a_B N_{\rm tot}$, and 
an ensemble of dust on the back side of, and 1.2~AU away from, the star would yield $N_{\rm tot} \sim3\times 10^{36}/a_B$.  
A rudimentary estimate, assuming an albedo of 4\% (cometary nuclei) thus gives a height not much less than the perimetrical 
dimension.   

If our hypothesis holds, that the same clump has been responsible for the yearly dip, the clump 
must be dynamically stable.  The mass we derived is only for the dust, and there is no evidence, even with a sufficient 
amount of associated gas, that the clump is on the verge of gravitational instability \citep{cha10}.  
In any case, the density of the 
clump is not likely to have a high contrast relative to the rest of the disk.  In other words, it may be just a density inhomogeneity, 
such as a local dust concentration in a warped, spiral-armed disk or density enhancement by a companion 
star \citep{gri98}, that gives rise to the characteristic light curve seen in GM\,Cep.

In conclusion, our photometric monitoring of GM\,Cep confirms its UXor nature.  Moreover, the light curves and color variations 
suggest density inhomogeneity of dust in the young stellar disk.  Such enhanced density contrast 
may be a signpost of the transition phase from grain growth to the onset of planetesimal formation.  
GM\,Cep may not be an isolated example, and intense monitoring should be carried out for young stars known to exhibit 
abrupt light variations.  Further characterization of the clumpy disk of GM\,Cep, e.g., by polarization, infrared 
spectroscopy, and high angular resolution submillimeter imaging, at epochs in and out of the occultation, should shed light 
on our hypothesis of this interesting young star.

\acknowledgments
We thank the anonymous referee for very constructive suggestions on an earlier version of the paper.
The work at NCU is financially supported in part by the grant NSC99-2119-M-008-021.
Lulin Observatory is operated by the National Central University of Taiwan, and is funded partially by the
National Science Council and Ministry of Education of Taiwan.  Jena co-authors thank DFG, the Thuringian government, 
the EU (MTKD-CT-2006-042514) for support. Jena and 
Torun co-authors thank DAAD PPP-MNiSW (50724260-2010/2011) for support.  Part of the observations reported here have been 
obtained with telescopes of the University Observatory Jena, which is operated by the Astrophysical Institute
of the Friedrich-Schiller-University

\end{document}